\documentclass[runningheads,a4paper]{llncs}
\usepackage{amssymb}
\usepackage{graphicx}
\usepackage{authblk}
\usepackage{blindtext}
\usepackage{listings}
\usepackage{amsmath}
\usepackage{subfig}
\usepackage{mwe}
\usepackage{lipsum}
\usepackage{float}
\usepackage{color}
\usepackage{csquotes}
\usepackage{hyperref}
\usepackage[colorinlistoftodos]{todonotes}

\usepackage{url}
\usepackage{booktabs}
\usepackage{footnote}

\newcommand\mymail[1]{\href{mailto:#1@iiitd.ac.in}{\nolinkurl{#1}}}

\newcommand\thanksrep[1][\value{footnote}]{\footnotemark[#1]}
\begin{document}
\mainmatter 

\title{White or Blue, the Whale gets its Vengeance: A Social Media Analysis of the Blue Whale Challenge}
\author{Abhinav Khattar\thanks{authors contributed equally} \and Karan Dabas\thanksrep \and Kshitij Gupta\thanksrep \and Shaan Chopra\thanksrep \and Ponnurangam Kumaraguru}
\pagestyle{plain}

\authorrunning{The Blue Whale Challenge: What, Why and How?}

\institute{IIIT Delhi, India\\\{\mymail{abhinav15120}, \mymail{karan15141}, \mymail{kshitij15048}, \mymail{shaan15090}, \mymail{pk}\}@iiitd.ac.in}
\maketitle
\begin{abstract}
The Blue Whale Challenge is a series of self-harm causing tasks that are propagated via online social media under the disguise of a \textquotedblleft game.\textquotedblright The list of tasks must be completed in a duration of 50 days and they cause both physical and mental harm to the player. The final task is to commit suicide. The game is supposed to be administered by people called \textquotedblleft curators \textquotedblright who incite others to cause self-mutilation and commit suicide. The curators and potential players are known to contact each other on social networking websites and the conversations between them are suspected to take place mainly via direct messages which are difficult to track. Though, in order to find curators, the players make public posts containing certain hashtags/keywords to catch their attention. Even though a lot of these social networks have moderated posts talking about the game, yet some posts manage to pass their filters. Our research focuses on (1) understanding the social media spread of the challenge, (2) spotting the behaviour of the people taking interest in Blue Whale challenge and, (3) analysing demographics of the users who may be involved in playing the game.

\end{abstract}

\section{Introduction}
The Blue Whale Challenge is organised in such a way so as to ultimately brainwash the minds of the players and drive them to cause self-harm. The tasks include waking up at odd hours, listening to psychedelic music, watching scary videos and inflicting cuts and wounds on their bodies~\cite{9}. This causes the players to have disturbed minds, making them more susceptible to influence.\textquotedblleft At some point, it is necessary to push the teenager not to sleep at night. [In this way, their] psyche becomes more susceptible to influence \textquotedblright, Philipp Budeikin said, explaining his tactics of manipulation~\cite{2}.
\begin{figure}
\centering
  \frame{\includegraphics[width=0.5\textwidth,keepaspectratio]{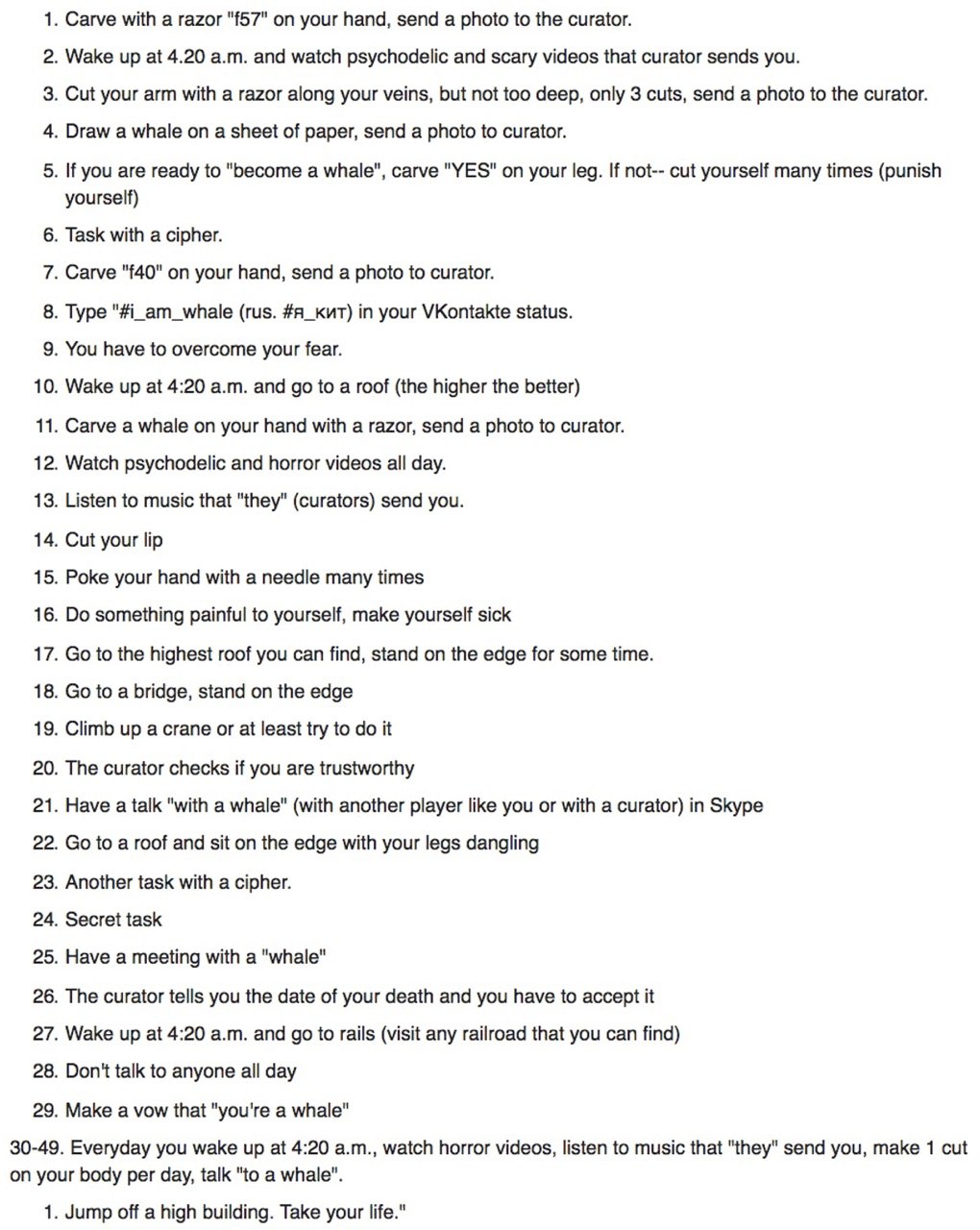}}
  \caption{The list of 50 Tasks as found on Reddit. We find some minor modifications in the list at different sources.}
  \label{fig:task}
\end{figure}

The curators and potential players use online social networks to contact each other. Curators seek out young people on social media who want to take part in the 50-day challenge and subject them to the tasks~\cite{26}. Also, a task of the game - Task 8 in Fig. \ref{fig:task} - asks the player to post to VKontakte (VK), a Russian social networking website. Posts of the challenge have now spread to  Twitter, Instagram, Facebook, Reddit and other social networks. 
The creator of the game, Philipp Budeikin, 21 years, was arrested for coaxing 16 schoolgirls to kill themselves~\cite{1}. Another Russian, Ilya Sidorov, 26 years, confessed to being the administrator of a so-called suicide group that had 32 under-age members~\cite{1}. The most recent administrator caught was a 17-year-old Russian girl who initially played the game but did not take her life in the end, instead turned into a curator~\cite{3}. A 17-year-old Chinese student was also arrested and charged with extremism over a Blue Whale chat group~\cite{19}. But even with the original curators in jail, how is the game still thriving? 
There are a lot of misconceptions about the game. Some people think that it exists as an APK file on playstore. Though there have been claims that at some point such applications existed on playstore, there is no such application as of now. People also think that it is a flash game available on some website. But the game actually thrives as a phenomenon on social media, not as an application. Some people are trying to hunt down public groups on social media websites like VK or Facebook~\cite{7}. A Chinese tech giant, Tencent, has found at least 12 groups on its QQ instant messaging service using keywords related to the Blue Whale game~\cite{18}. Other reports claim that the game is fake and made the news because of spread of misinformation~\cite{6}. Several cases of suicide and self-harm have not gained as much \textquotedblleft popularity\textquotedblright as compared to similar incidents that are said to be caused as a result of the challenge and it is difficult to point out suicides that were caused solely due to the game~\cite{5}. \textquotedblleft It is said that reckless journalism actually created a sense of panic about the Blue Whale challenge when the real concern should actually be addressing mental illness.\textquotedblright~\cite{6} 
There are also claims that accounts of people are hacked and used as \textquotedblleft curator\textquotedblright accounts to incite others and that at some point in time, there were links on Facebook leading to the game~\cite{8}. Players were previously sought out through “death groups” on social networks like VK. It is believed that members of these groups contacted curators via direct message and primarily conversed in Russian~\cite{7}. \textquotedblleft So, if the game's curators are in prison, how are teens still playing it? Well, the problem is, with all these teens out there searching for the game, all they need to do is stumble across one individual who is looking to exploit them and very soon they'll be playing Blue Whale, or at least a copy cat's version of the game\textquotedblright~\cite{7}. Any person on the internet can claim to be a curator and continue the game, inciting more and more people to commit self-harm. They can also be bots that act as curators and send messages to people.  Since the primary link between victims and curator(s) is using online social media, we can try to find out users who might be vulnerable and spot common properties these accounts might share. Governments and authorities across the world are trying to take steps to curb the spread of the Blue Whale Challenge. 
We wish to (1) understand the social media spread of the challenge, (2) spot the behaviour of the people taking interest in Blue Whale challenge and, (3) analyse demographics of the users who may be playing the game. 

\section{Related Work}
Scientific literature was reviewed to get a deeper understanding of the challenge and how it thrives online. In~\cite{16}, the authors studied the structure of the tasks to understand how the challenge brainwashes the players who take it. They further explain how the instructions of the game exploit fear psychology to prepare the victims for self-infliction of pain and suicide: Tasks 1 to 9 serve as the induction tasks followed by the habituation tasks - 10 to 25 - followed by the final preparation tasks - 26 to 50. Also, the authors found that teenagers with complicated upbringing and negative life experiences are more likely to get involved in the game. Another paper~\cite{17} by Jouni Smed et al assesses the negative effects of gamification such as game addiction and ethical issues. The Blue Whale challenge shows how gamification can be used for harmful purposes, that is used to engage the users and drive them towards suicide. 

\section{Methodology} 
\begin{figure}
\centering
  \includegraphics[width=0.7\textwidth,keepaspectratio]{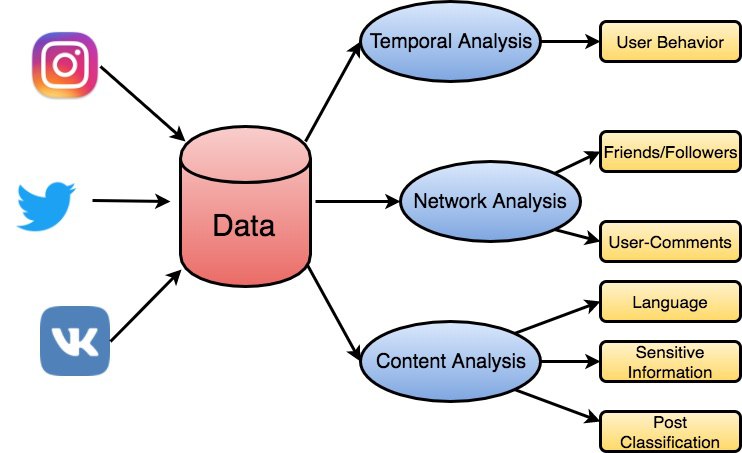}
\caption{Architecture Diagram for Methodology}
  \label{arch}
\end{figure}

Data was collected from three social media websites: VK, Instagram, and Twitter. The aforementioned Social Networks were chosen for the analysis as data was readily available on them. Table \ref{Table:1} shows the duration for which posts from social networks were collected. Fig. \ref{arch} shows the architecture diagram for the methodology followed for the study.

\begin{savenotes}
\begin{table}
\centering
\caption{Duration for which data from different social networks was collected(GMT)}
\label{Table:1}
\begin{tabular}{@{}lllll@{}}
\toprule
Social Media && \#First Post Date(dd-mm-yyyy)  && \#Last Date for Data Collection(dd-mm-yyyy)           \\ \midrule
VK           && 01-03-2017     && 01-10-2017                                \\
Instagram    && 03-07-2013     && 01-10-2017                               \\
Twitter          && 18-08-2017     && 01-10-2017                                \\ \bottomrule
\end{tabular}
\end{table}
\end{savenotes}
We collected data for the following Hashtags: \#i\_am\_whale, \#curatorfindme, \#f57, \#wakemeupat420, \#iamawhale, \#I\_am\_whale, \#iamwhale, \#imwhale. Initially, the collected posts contained either of the two hashtags: \#curatorfindme or \#i\_am\_whale. Query expansion was used to include the following hashtags in our analysis as well: \#f57, \#wakemeupat420, \#iamawhale, \#I\_am\_whale, \#iamwhale, \#imwhale. Hashtags like ‘\#bluewhalechallenge’ and ‘\#bluewhale’ were avoided as they contained a lot of noise (posts which were news related or spreading awareness about the challenge). We performed various analysis on the collected data, categorising it into Temporal Analysis, Content Analysis, and Network Analysis.

\section{Statistics}
\begin{figure}
\centering
  \includegraphics[width=0.5\textwidth,keepaspectratio]{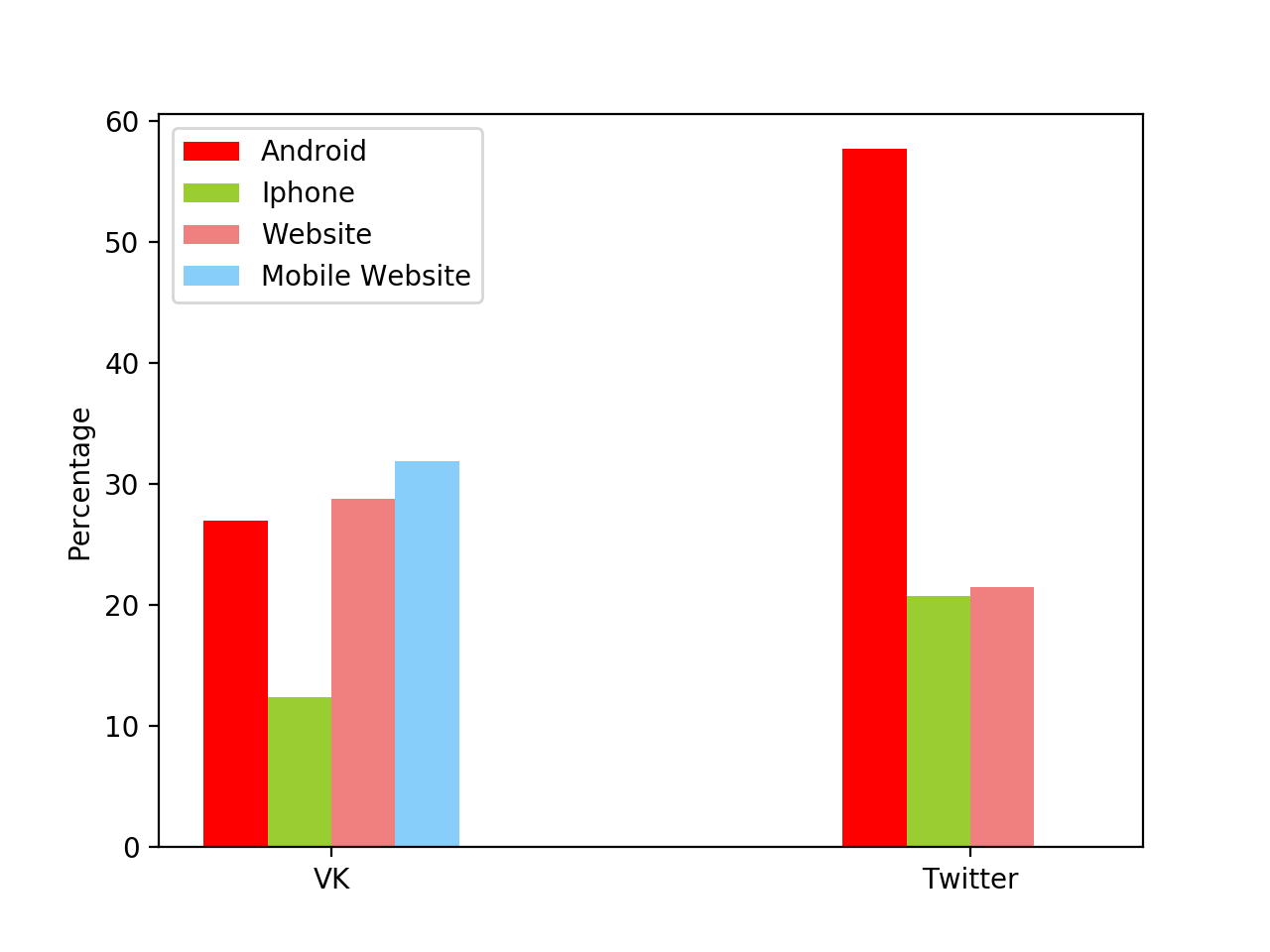}
  \caption{Platforms used to access various social media sites and post about the Blue Whale Challenge.}
  \label{fig:2}
\end{figure}

\begin{figure}
\centering
  \frame{\includegraphics[width=0.3\textwidth,keepaspectratio]{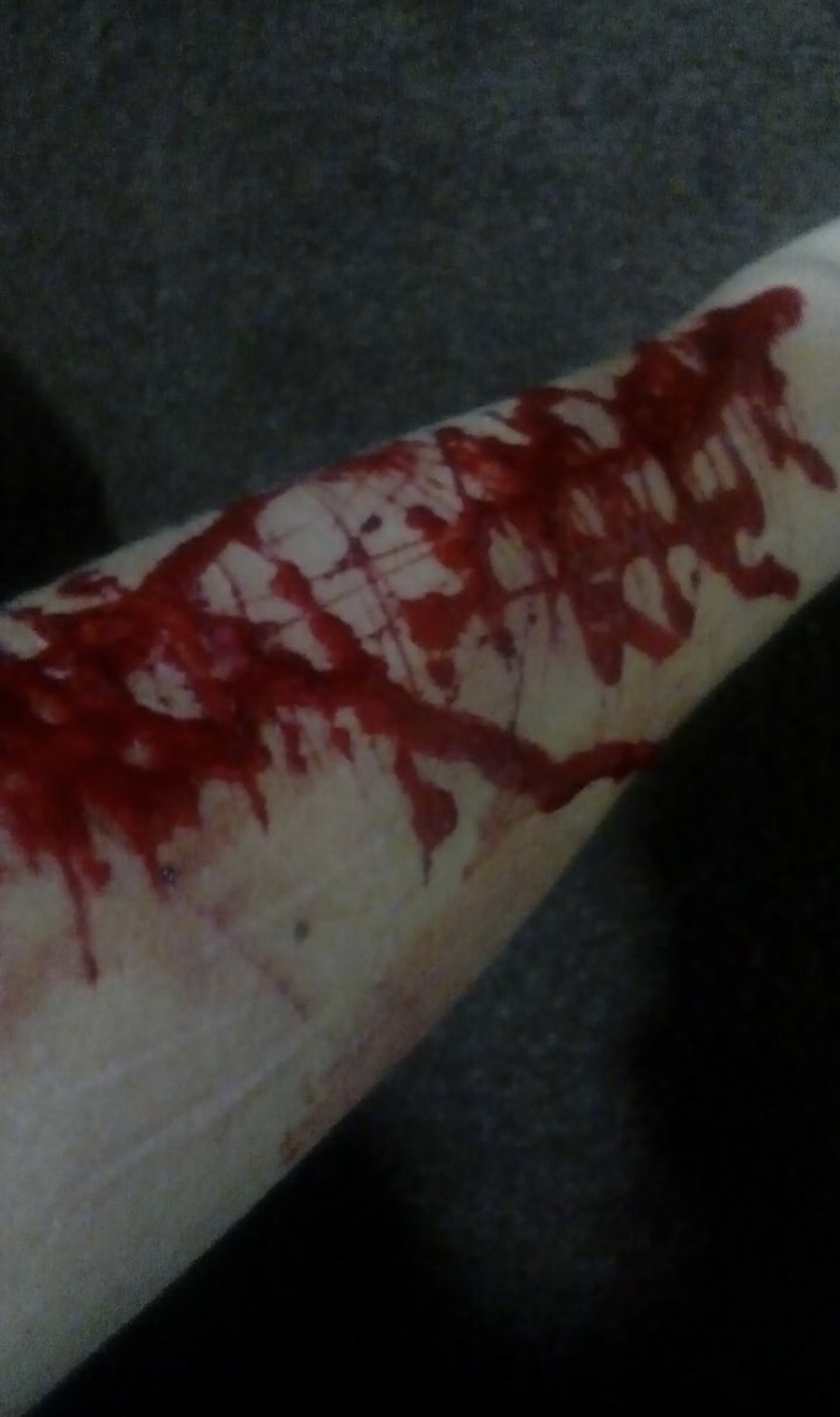}}
\caption{The image posted on VK - initially collected as a part of our dataset - was deleted. Interestingly, the entire user account has been temporarily suspended. The text content of the post was \#i\_am\_whale.}
  \label{susp}
\end{figure}

Table \ref{Table:2} shows the number of cases associated with the Blue Whale Challenge - including those who died, who were saved and who showed signs of playing the game. We referred to various articles from news sources; except for one news source from Chile with highest global Alexa ranking of 130,040, the ranking of all other news sources fell under 45,000. The news source from Chile was a regional newspaper with country-wise Alexa score of 721. Finding the exact number of deaths due to this is extremely difficult. According to news sources, around 130 deaths are associated with the Blue Whale Challenge in Russia and all these teenagers were known to be part of the same internet group~\cite{24,23}. According to the Russian investigation though, only 8 deaths were actually due to the Blue Whale challenge~\cite{25}. India ranks highest according to Google trends in terms of searches related to the Blue Whale game in 12 months~\cite{13,11}. Table \ref{Table:3} captures the various demographics of the data we collected. Fig. \ref{fig:2} shows the devices used to post about the challenge on VK and Twitter. We can see that a lot of posts on VK have been made using the mobile website; this can be because the Russian social network was recently banned in India due to the challenge~\cite{20}. When we cross-checked the collected posts on 13th October 2017, a number of them had already been deleted. Fig. \ref{susp} shows the image contained in such a deleted post on VK. Table \ref{Table:3} shows these figures for different social networks. Comments on Twitter stands for replies.

\begin{figure}
\centering
  \includegraphics[width=0.6\textwidth,keepaspectratio]{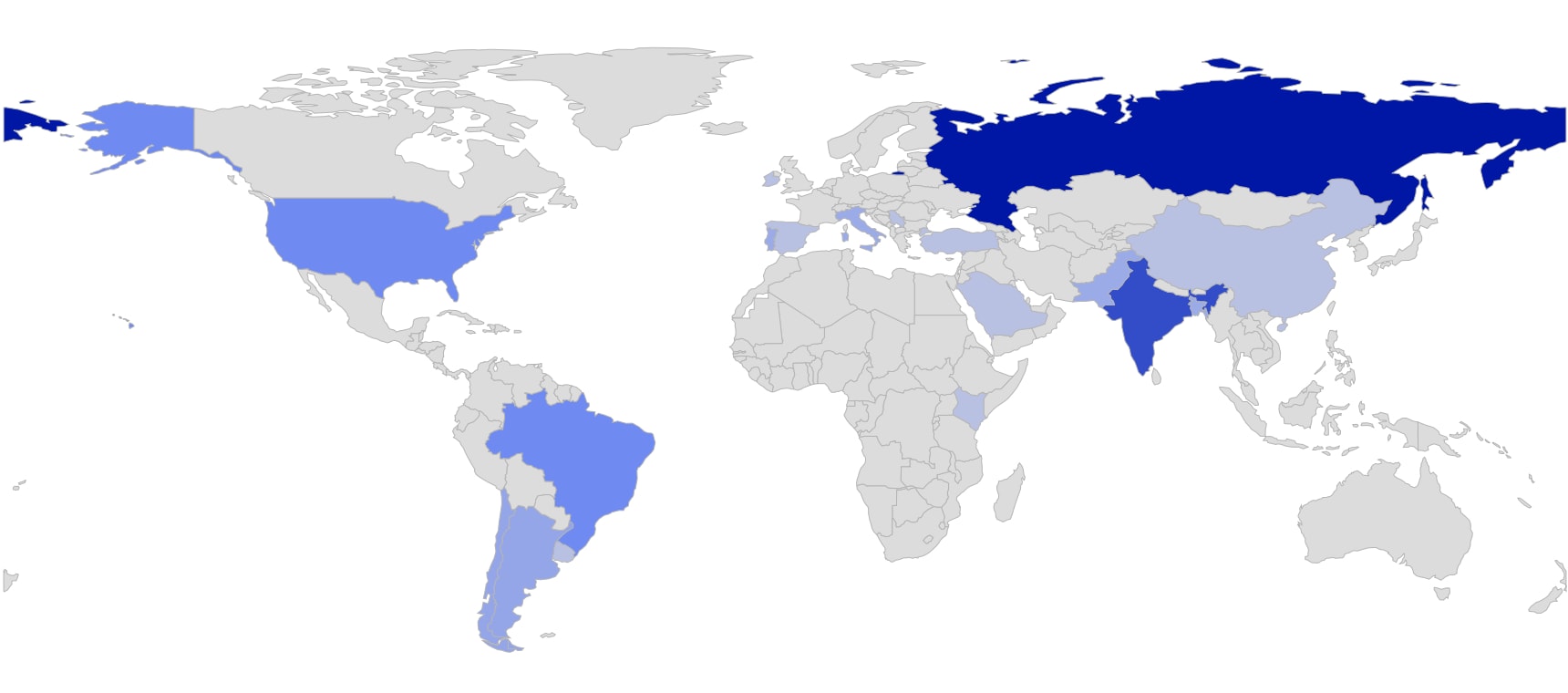}
  \caption{Distribution of Blue Whale cases across the world}
  \label{carto}
\end{figure}

\begin{savenotes}
\begin{table}
\centering
\caption{Number of cases related to the Blue Whale Challenge in different countries}
\label{Table:2}
\begin{tabular}{@{}llll@{}}
\toprule
Country & \#Cases & \#Country & \#Cases           \\ \midrule
Argentina           & 3 & Pakistan          & 2  \\
Bangladesh    & 2 & Portugal           & 2        \\
Brazil          & 4  & Russia           & 130 \footnote{Around 130 suicides in Russia are linked to the Blue Whale Challenge~\cite{24,23}} \\
Chile          & 3 & Saudi Arabia           & 1 \\
China          & 1  & Serbia           & 1 \\
India          & 10  & Spain           & 1 \\
Ireland & 1   & Turkey           & 1 \\
Italy          & 2 &  United States    & 4 \\
Kenya         & 1 & Uruguay           & 1 \\ \midrule
&   &		TOTAL & 170 \\
\bottomrule
\end{tabular}
\end{table}
\end{savenotes}

\begin{table}
\centering
\caption{Data Description}
\label{Table:3}
\begin{tabular}{@{}lllll@{}}
\toprule
Social Media & \#Posts & \#Unique users & \#Comments & \#Deleted posts            \\ \midrule
VK           & 862     & 705           & 894       & 76                      \\
Instagram    & 1,137     & 736           & 751       & 386                       \\
Twitter          & 677     & 548           & 27       & 83                       \\ \bottomrule
\end{tabular}
\end{table}


\section{Analysis}
We divide our analysis into 3 broad categories: Temporal, Content, and Network; with graphs from each of the 3 networks (where available).

\subsection{Temporal Analysis}
\begin{figure}[!htb]
\centering
\subfloat[][VK]{\includegraphics[width=0.55\textwidth,keepaspectratio]{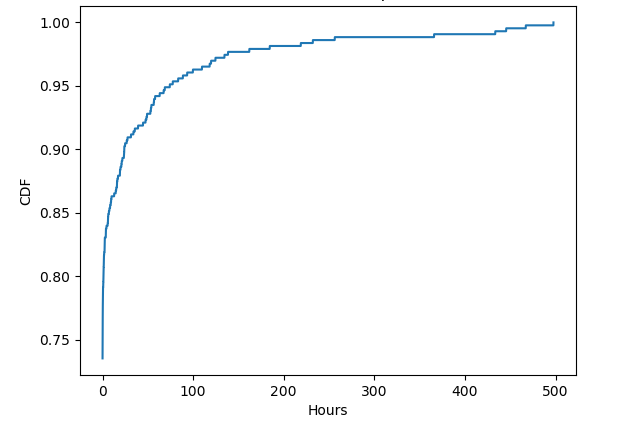}}
\centering
\subfloat[][Instagram]{\includegraphics[width=0.55\textwidth,keepaspectratio]{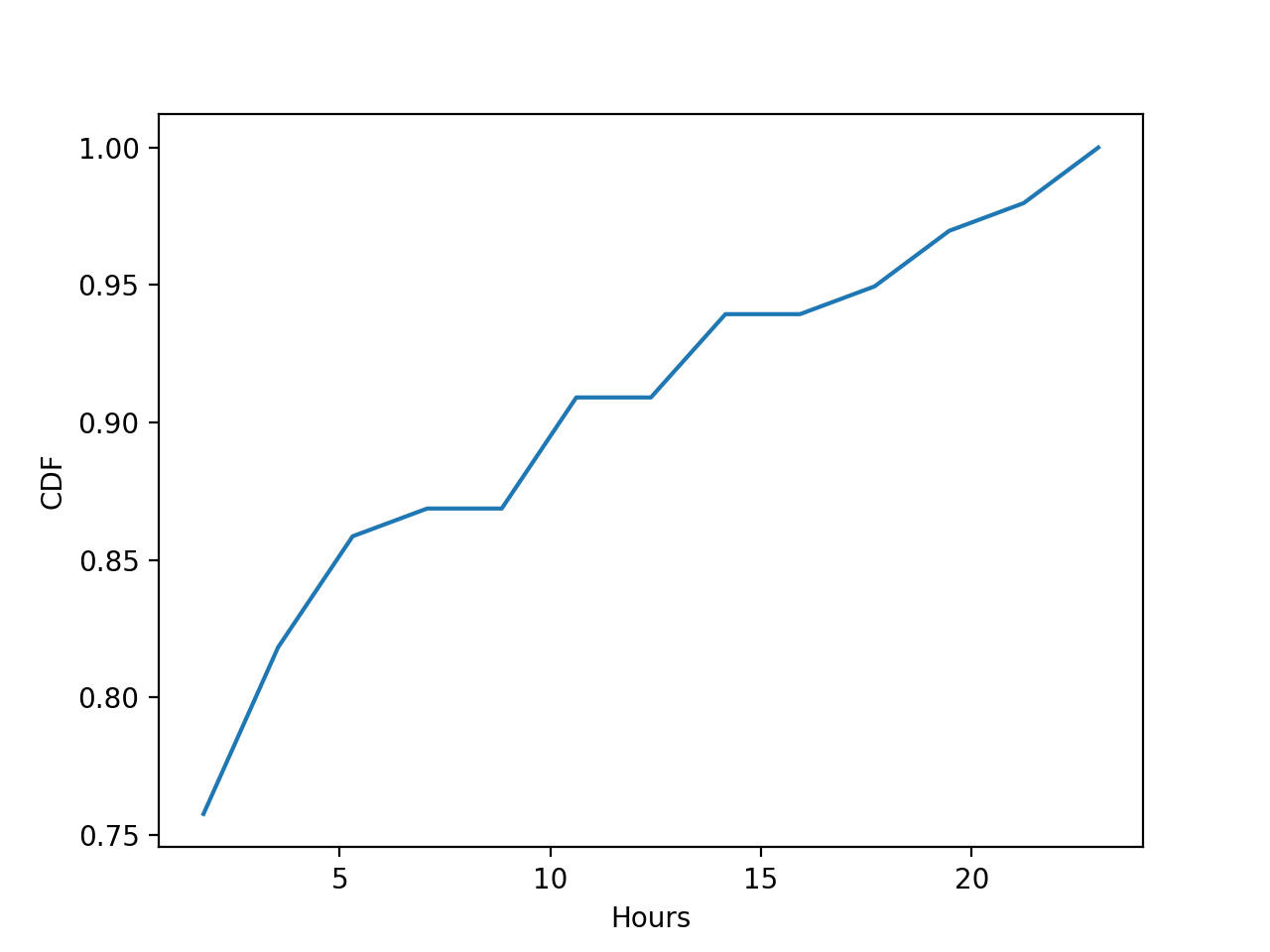}}\\
\centering
\subfloat[][Twitter]{\includegraphics[width=0.55\textwidth,keepaspectratio]{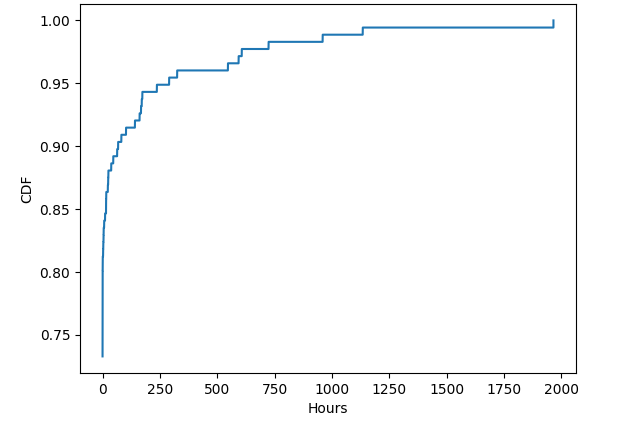}}
\centering
\caption{Time difference between first and last posts related to Blue Whale on different social networks}
\label{fig:3}
\end{figure}

\begin{figure}[!htb]
\centering
\subfloat[][Indegree - number of followers - of \\users posting about the Blue Whale \\Challenge]{\includegraphics[width=0.5\linewidth,keepaspectratio]{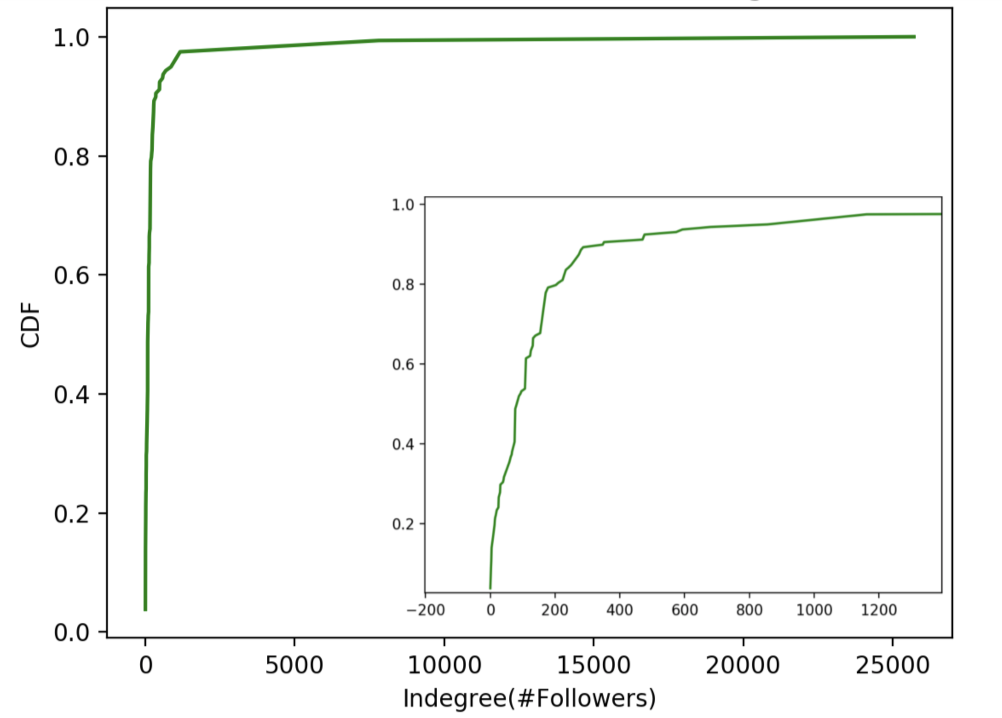}}
\centering
\subfloat[][Outdegree - number of followings - of users posting about the Blue Whale \\Challenge]{\includegraphics[width=0.5\linewidth,keepaspectratio]{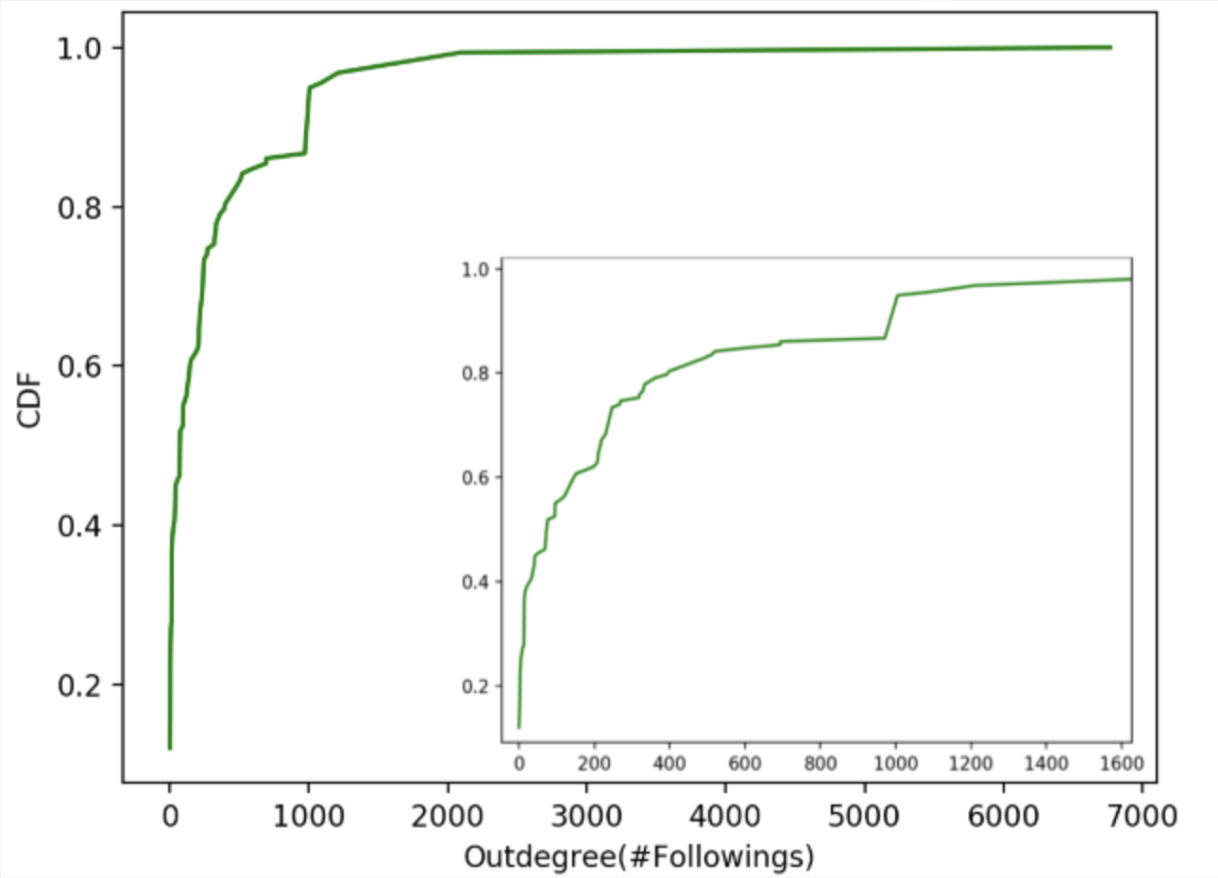}}
\caption{Instagram}
\label{fig:4}
\end{figure}

\begin{figure}[!htb]
\centering
  \includegraphics[width=0.6\textwidth,keepaspectratio]{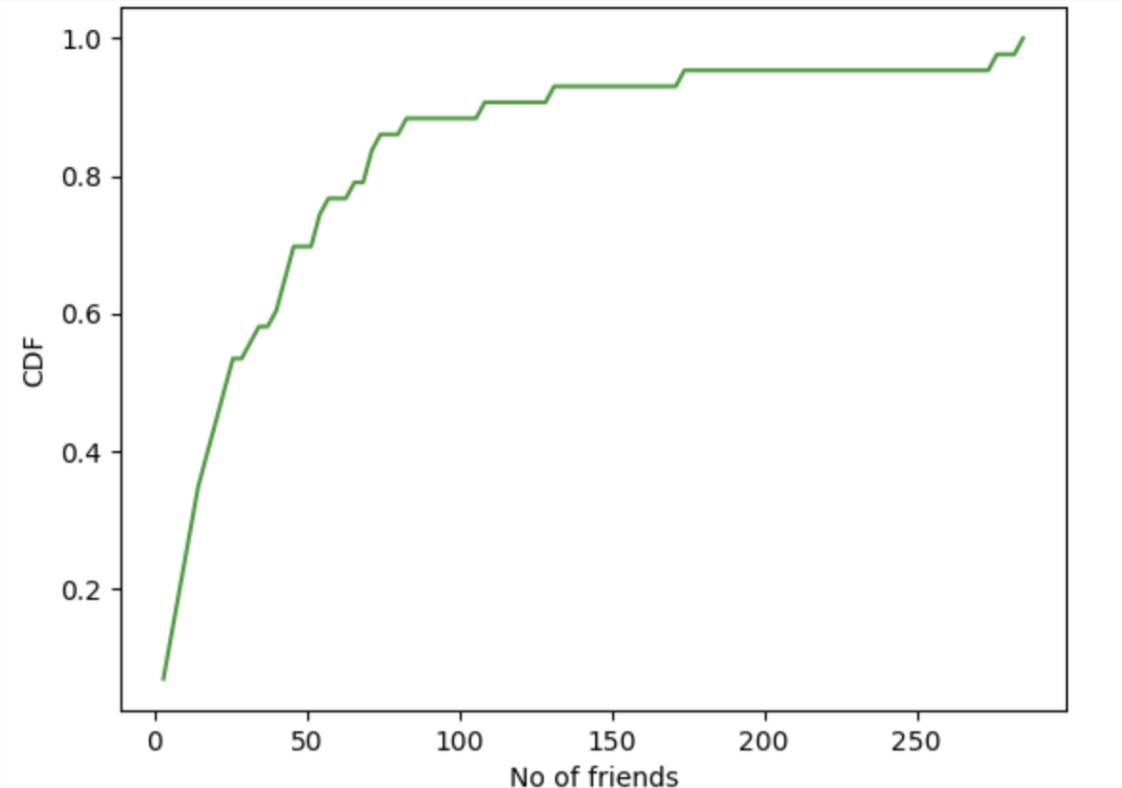}
  \caption{VK - Total number of Blue Whale posts with respect to the number of friends of the users}
  \label{fig:5}
\end{figure}
In general, we observe that people on VK and Twitter continued to post about the Blue Whale Challenge/followed the challenge even multiple days after their initial post. We deemed a post Blue Whale related if the text contained any of the relevant hashtags as used in the data collection. Most of the content in Blue Whale related posts was hashtags. The statistics from Instagram show a really low follow-up time. Though Instagram isn't removing all the posts, but instead, if any of the sensitive hashtags are searched for, they ask the user whether they need support but still give an option to see the posts anyway. In Fig. \ref{fig:3}(a), it is seen that the time difference between the first and last post related to Blue Whale on VK by users ranges from 0 hours to 500 hours. Around 85\% of the users continued posting about Blue Whale for less than an hour. 98.6\% of the users have a time difference of fewer than 200 hours between their first and last Blue Whale related post. In Fig. \ref{fig:3}(b), it is seen that the time difference between the first and last post related to Blue Whale on Instagram by users ranges from 0 hours to 22.5 hours. 94\% of the users have a time difference less than 15 hours between their first and last Blue Whale related post. In Fig. \ref{fig:3}(c), it is seen that the time difference between the first and last post related to Blue Whale on Twitter goes up to 2,000 hours. But similar to VK, 81.25\% of the users continued posting about the challenge for less than an hour. \\
In Fig. \ref{fig:4}(a), it is seen that around 70\% of users talking about Blue Whale challenge on Instagram have less than 200 followers. In Fig. \ref{fig:4}(b), it is seen that around 60\% of users talking about Blue Whale challenge on Instagram have less than 200 followings. In Fig. \ref{fig:5}, it is seen that 60\% of the users talking about the Blue Whale Challenge on VK have up to 50 friends. We observe that most of the users who posted about the Blue Whale Challenge on both VK and Instagram did not have a high number of Followers/Friends. On manual verification, we also found that most of these user IDs were actually new and only contained posts about the challenge. On Twitter, 27.01\% of the user accounts in our dataset were created in 2017.

\subsection{Network Analysis}
\begin{figure}
\centering
  \includegraphics[width=0.6\textwidth,keepaspectratio]{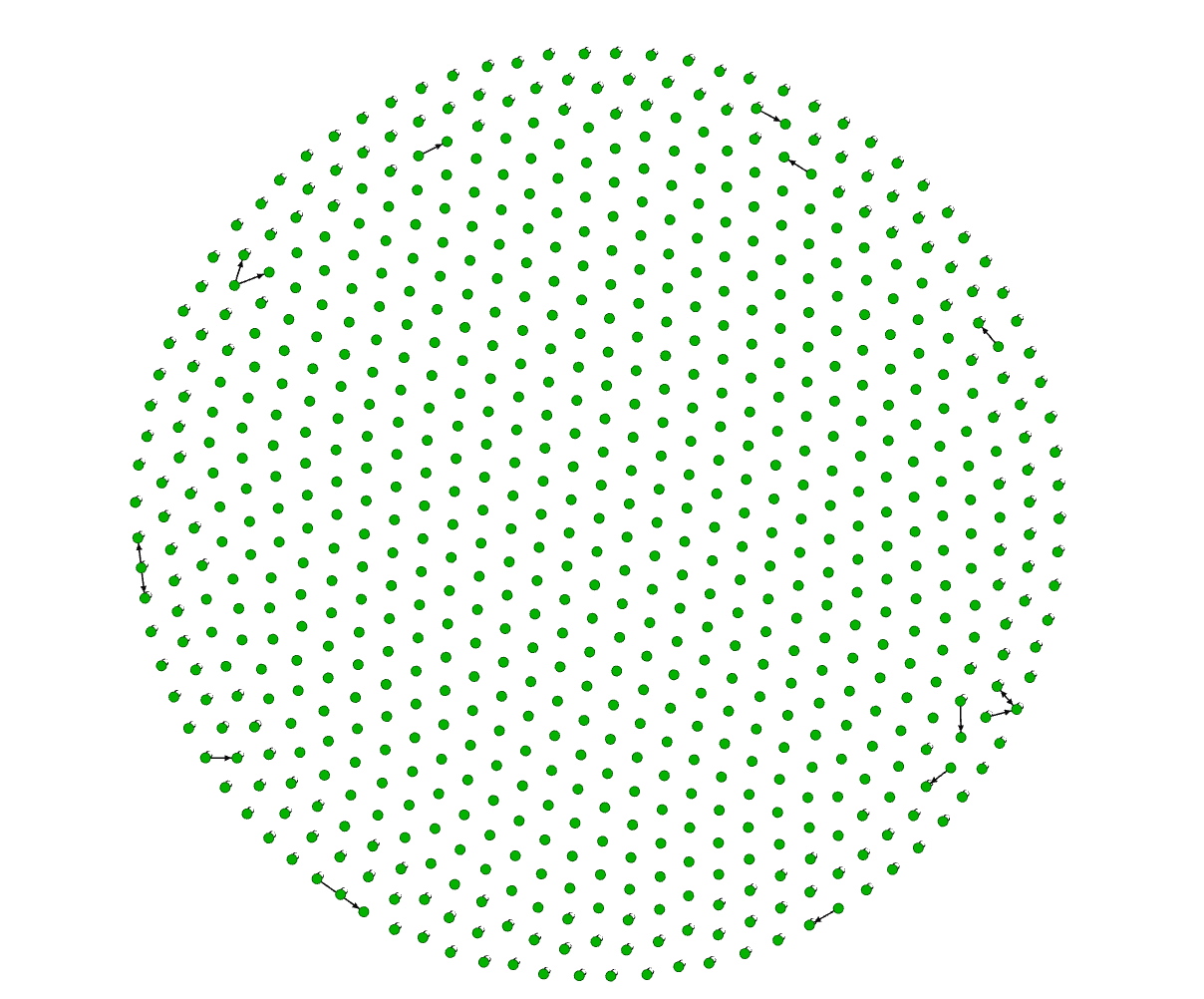}
  \caption{Instagram Comments Graph: If a user comments on the post by another user, there is a directed edge between them}
  \label{fig:6}
\end{figure}

\begin{figure}[!htb]
\centering
\subfloat[][VK Friends Graph: If 2 users are friends then there \\is an edge between them]{\includegraphics[width=0.55\textwidth,keepaspectratio]{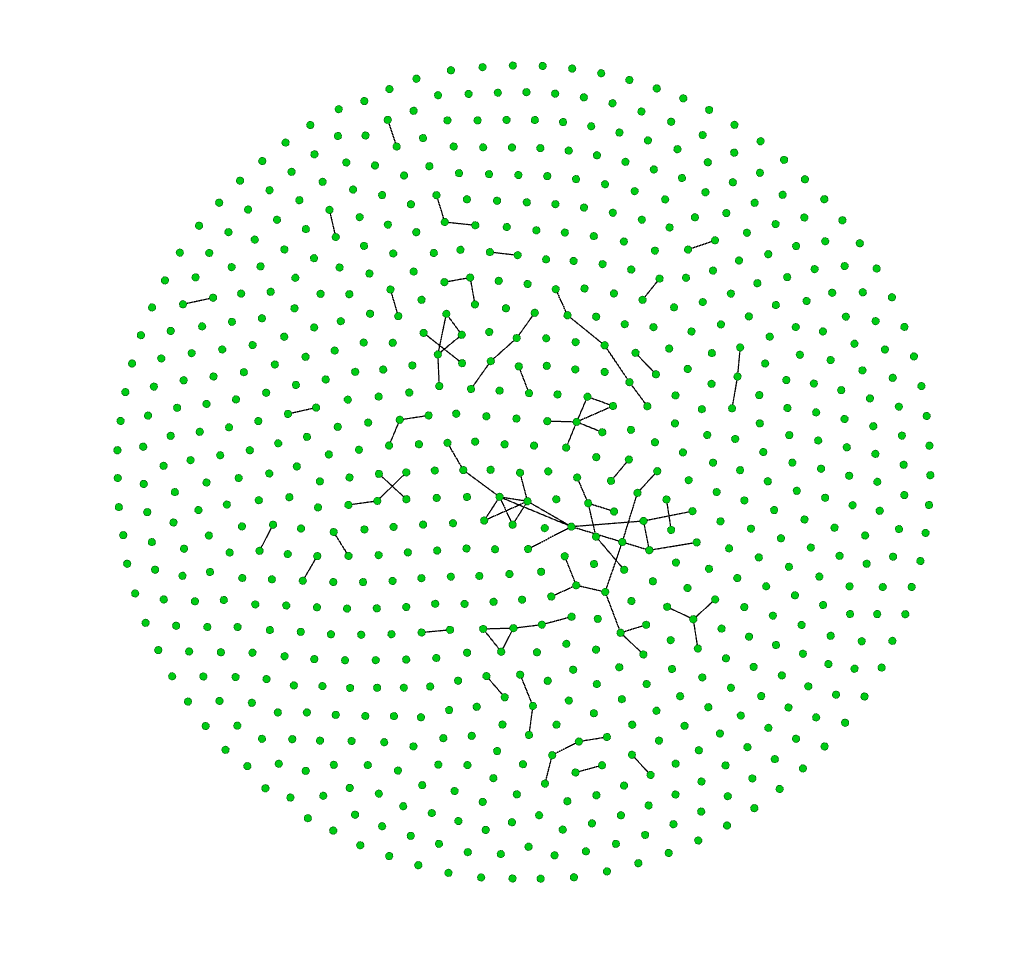}}
\centering
\subfloat[][VK Comments Graph: If a user comments on the post by another user, there is a directed edged between them]{\includegraphics[width=0.55\textwidth,keepaspectratio]{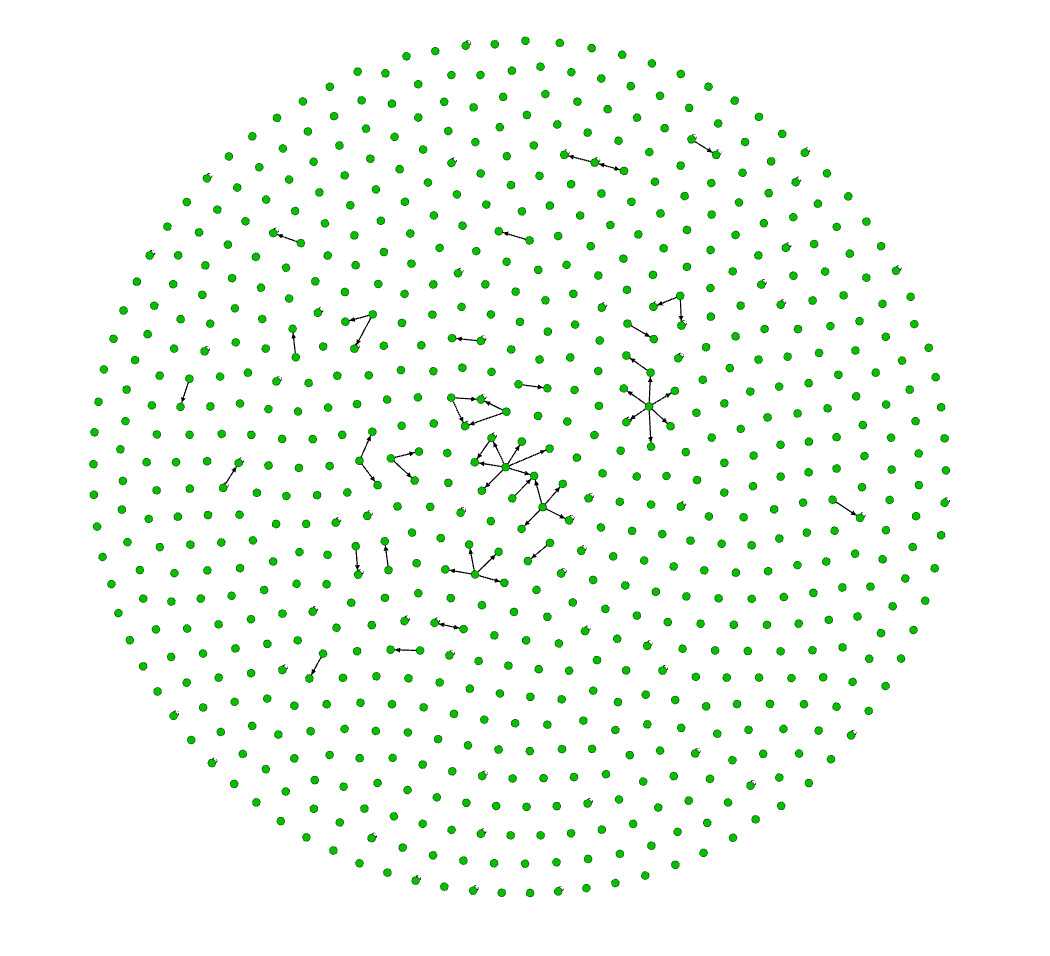}}
\caption{VK - Network Analysis}
\label{fig:7}
\end{figure}

Fig. \ref{fig:6} and Fig. \ref{fig:7} depict the network of the users we got from Instagram and VK respectively. Fig. \ref{fig:7}(a) shows links between users that are friends on VK and are present in the data collected. Fig. \ref{fig:6} and Fig. \ref{fig:7}(b) show links between users based on their comments on others' posts, that is, there is an edge from A to B if A commented on B's post. As we can see the comment network for Instagram is much more sparse as compared to VK. The comment network on VK has an average clustering coefficient of 0.012. From Fig. \ref{fig:7}(a), we also observe that certain users posting about the Blue Whale Challenge are inter-connected on VK, that is, most of the users in this subset tend to be friends on VK and form communities. The average clustering coefficient of the VK friends graph comes out to be 0.262 - this excludes the nodes which have no edges. On the other hand, we were not able to find any follower-following link amongst the users present on Instagram.

\subsection{Content Analysis}
\subsubsection{Language Analysis:}
\begin{figure}
\centering
\subfloat[Instagram]{\includegraphics[scale=.12]{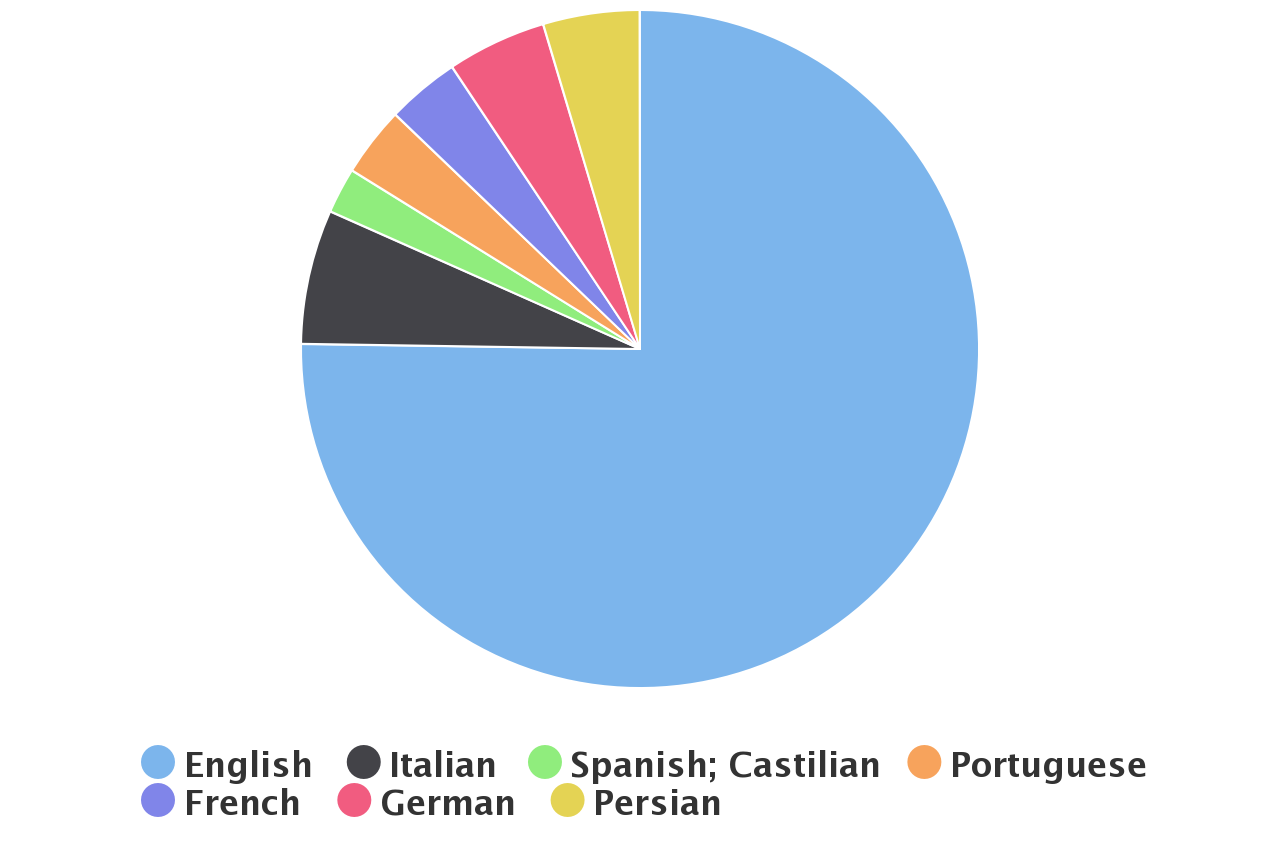}}
\subfloat[Twitter]{\includegraphics[scale=.12]{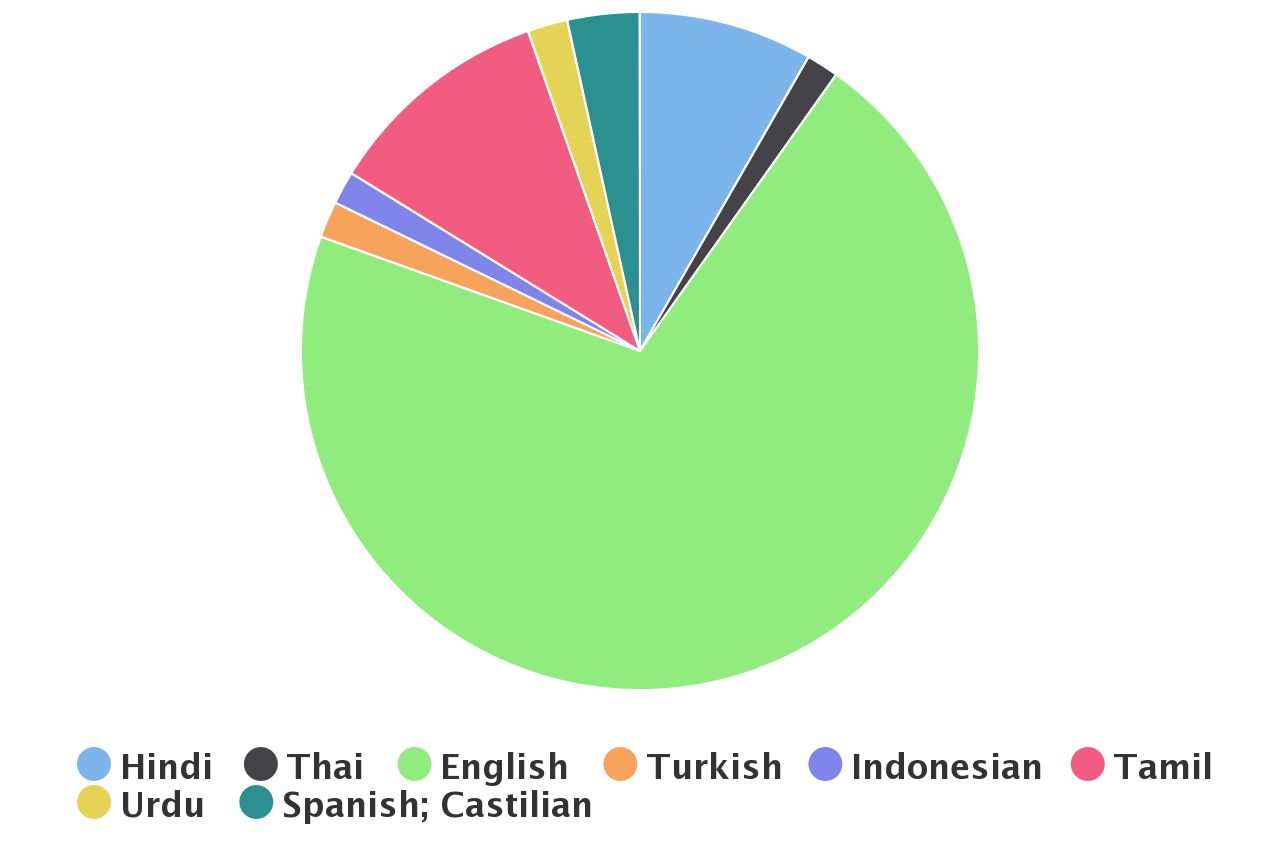}} \\
\subfloat[VK]{\includegraphics[scale=.12]{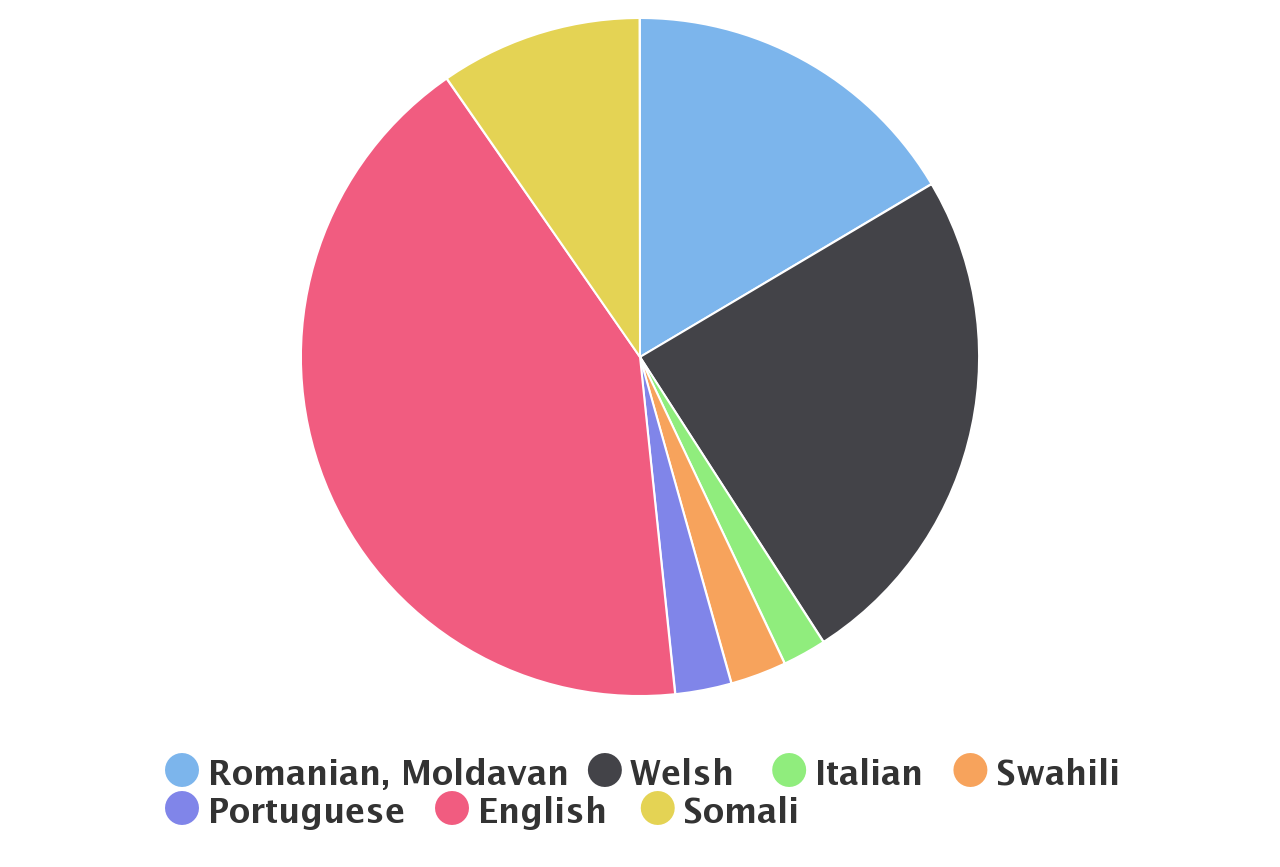}}

\caption{Different languages on different social networks in which Blue Whale related posts are made}
\label{fig:lang}
\end{figure}

We used a port of Google's language detection library in python - \href{https://pypi.python.org/pypi/langdetect}{langdetect} - to determine the languages of the posts. English is the most commonly used language in the three social networks.
In Fig. \ref{fig:lang}(a) it is seen that Italian, Persian and German are used in almost equal number of posts on Instagram. In Fig. \ref{fig:lang}(b) it is seen that Tamil and Hindi are used in quite a few posts on Twitter. In Fig. \ref{fig:lang}(c) it is seen that a good number of posts on VK are made in Welsh, Romanian and Somali languages.

\subsubsection{Sensitive Information:}
\begin{figure}
\centering
  \includegraphics[width=0.5\textwidth,keepaspectratio]{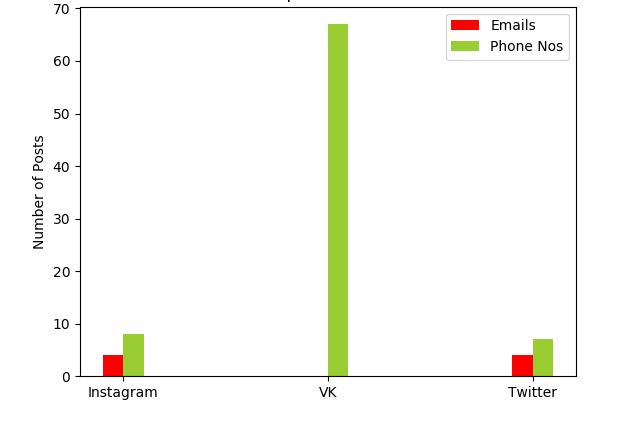}
  \caption{Amount and type of sensitive information shared across various social media sites to play Blue Whale game}
  \label{fig:sensitive}
\end{figure}

People reveal sensitive information about themselves like their email addresses and phone numbers so that curators can contact them. In Fig. \ref{fig:sensitive}, we see that around 70 phone numbers are revealed by users in posts and comments about the Blue Whale challenge on VK. Some email addresses and phone numbers are also revealed by users on Twitter and Instagram.

\subsubsection{User Mentions}
\begin{figure}
\centering
\subfloat[Account that posted images of cut legs]{\frame{\includegraphics[scale=.4]{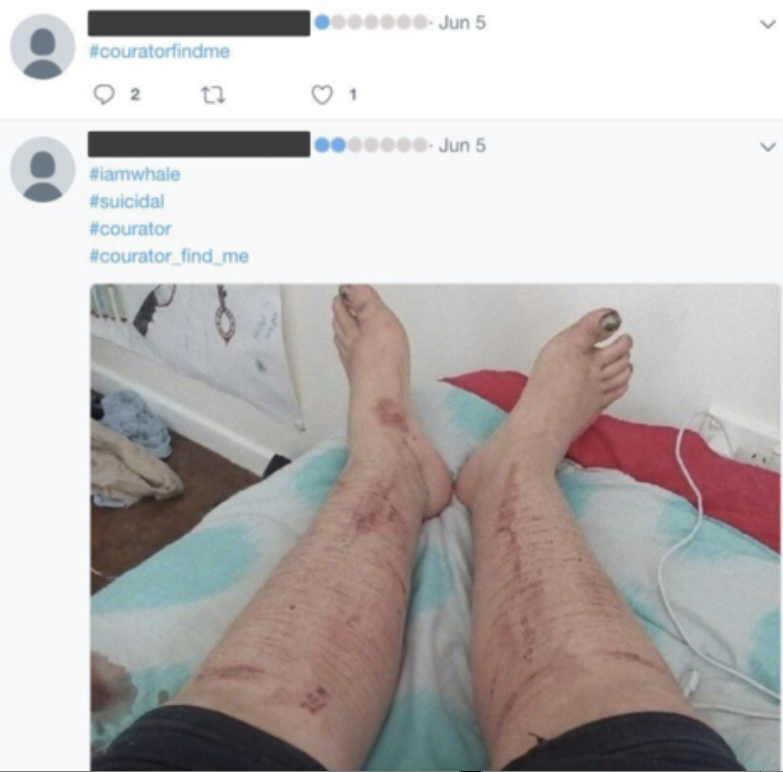}}}
\hfill
\subfloat[Account that posted images of cut wrist]{\frame{\includegraphics[scale=.4]{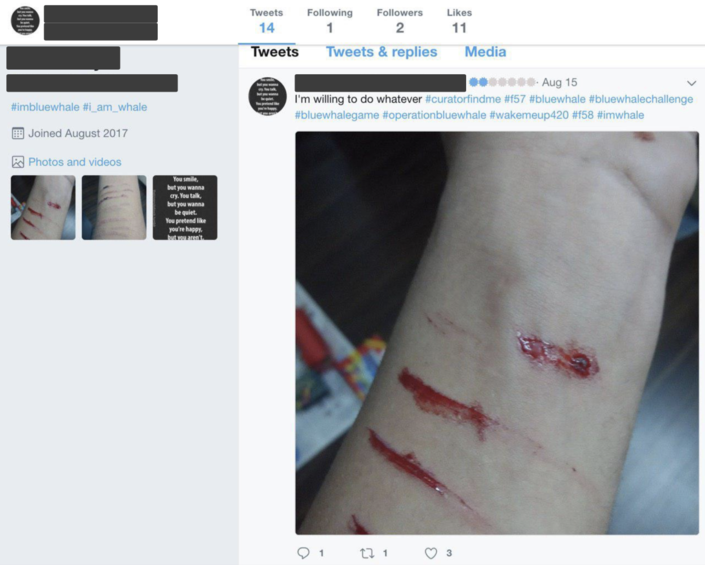}}} \\
\centering
\subfloat[Account that used popular Blue Whale related hashtags to catch attention]{\frame{\includegraphics[scale=.4]{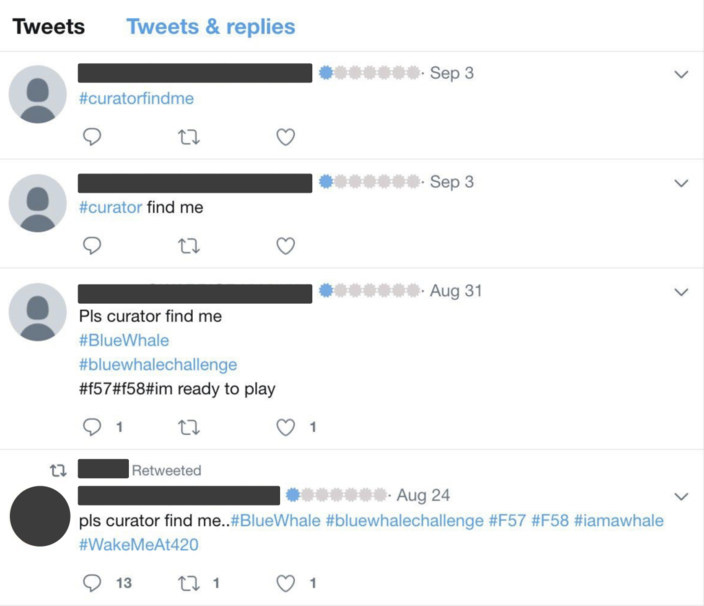}}}

\caption{Examples of some typical behaviours showed by user mentioned accounts on Twitter}
\label{userMention}
\end{figure}

\begin{figure}
\centering
  \frame{\includegraphics[width=0.5\textwidth,keepaspectratio]{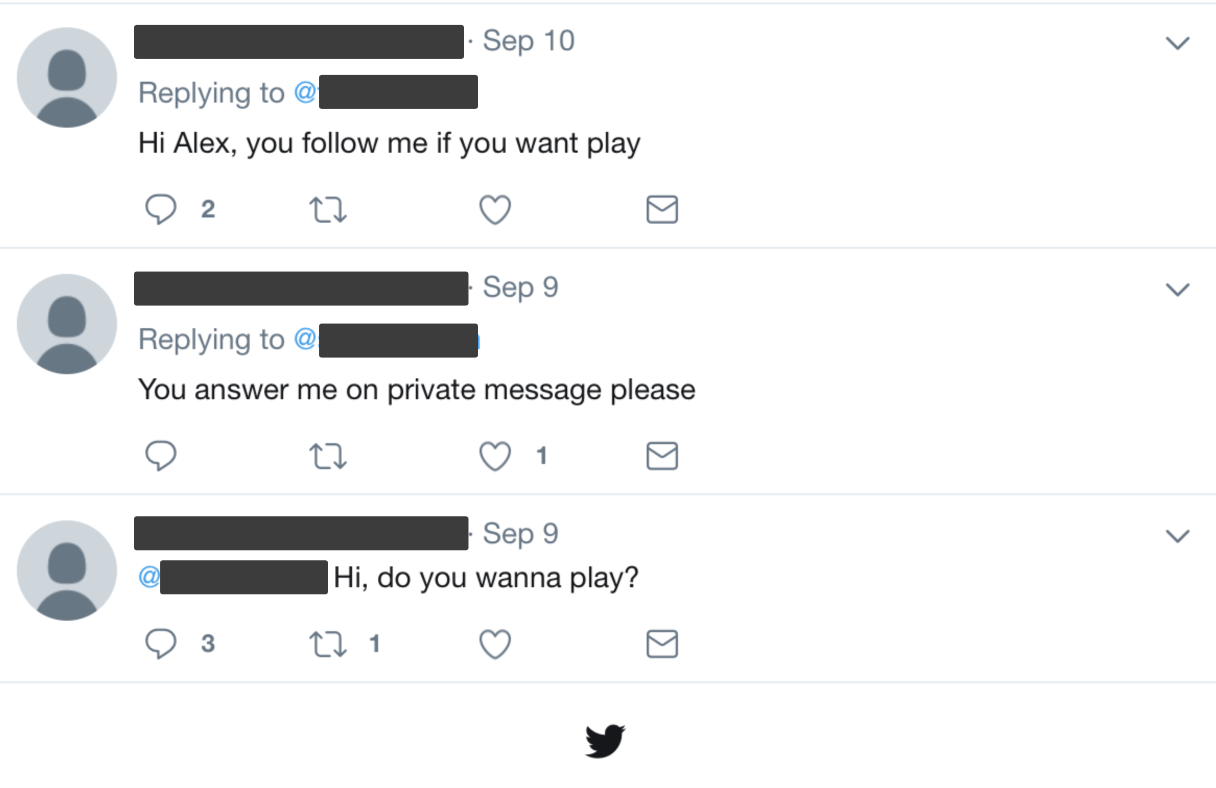}}
  \caption{A user-mentioned account on Twitter pretending to be a curator}
  \label{fig:curator}
\end{figure}

144 unique user accounts were mentioned in the collected twitter dataset. Most of these user-mentioned accounts are either famous users or users who are trying to stop this game; the list also contains accounts of Twitter security and Twitter. But there were a few accounts that tweeted only about the Blue Whale challenge. Some typical behaviours shown by these user accounts are: (1) tweets containing less text and more hashtags so as to catch the attention of users including curators, (2) occasional pictures of results of tasks such as carved arms and legs, and (3) low follow-up, that is, most users don't follow up after 1-3 posts about the challenge. Fig. \ref{userMention} shows these behaviours by accounts found as user-mentions in the collected tweets. \\
We also came across an interesting account that was acting to be a curator and asking users to private message or follow him/her if they wanted to join the game. Fig. \ref{fig:curator} shows the profile of this Twitter account.

\section{Different types of Users involved in the game}

\subsection{Potential Victims}
\begin{figure}
\centering
  \frame{\includegraphics[width=0.6\textwidth,keepaspectratio]{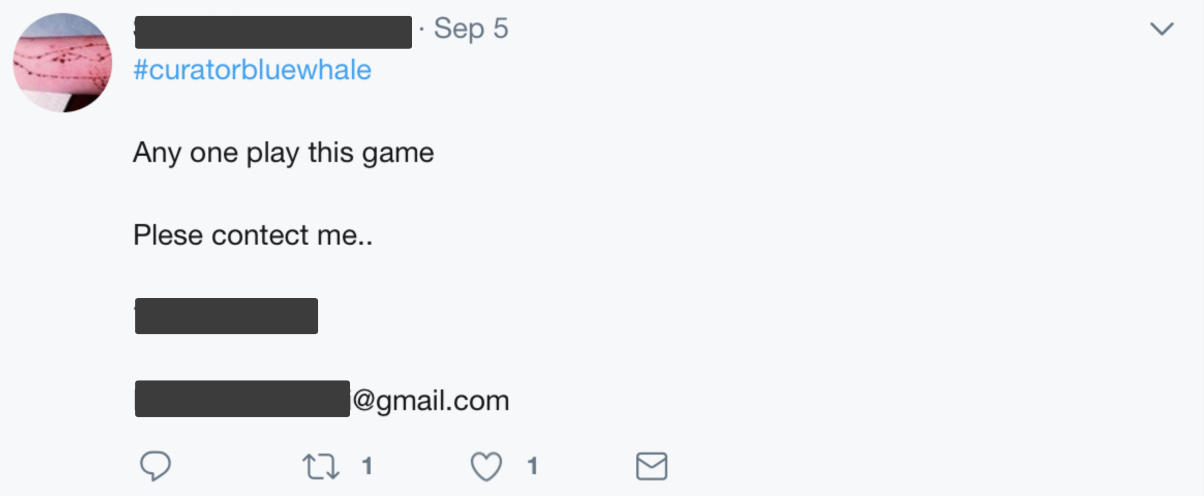}}
  \caption{Twitter - A post containing contact information revealed by a user in pursuit of joining the game }
  \label{twitter_post}
\end{figure}
Users who are depressed and ready to go to any extent to become a part of the game fall under this category. Such users often tend to reveal personal information like phone numbers, email addresses etc. so that curators can contact them. Fig. \ref{twitter_post} shows a tweet where a user revealed his/her contact information.

\begin{figure}

\centering
\subfloat[A WhatsApp Group specifically made for people willing to be a part of the Blue Whale Challenge]{\label{main:a}\frame{\includegraphics[scale=.3]{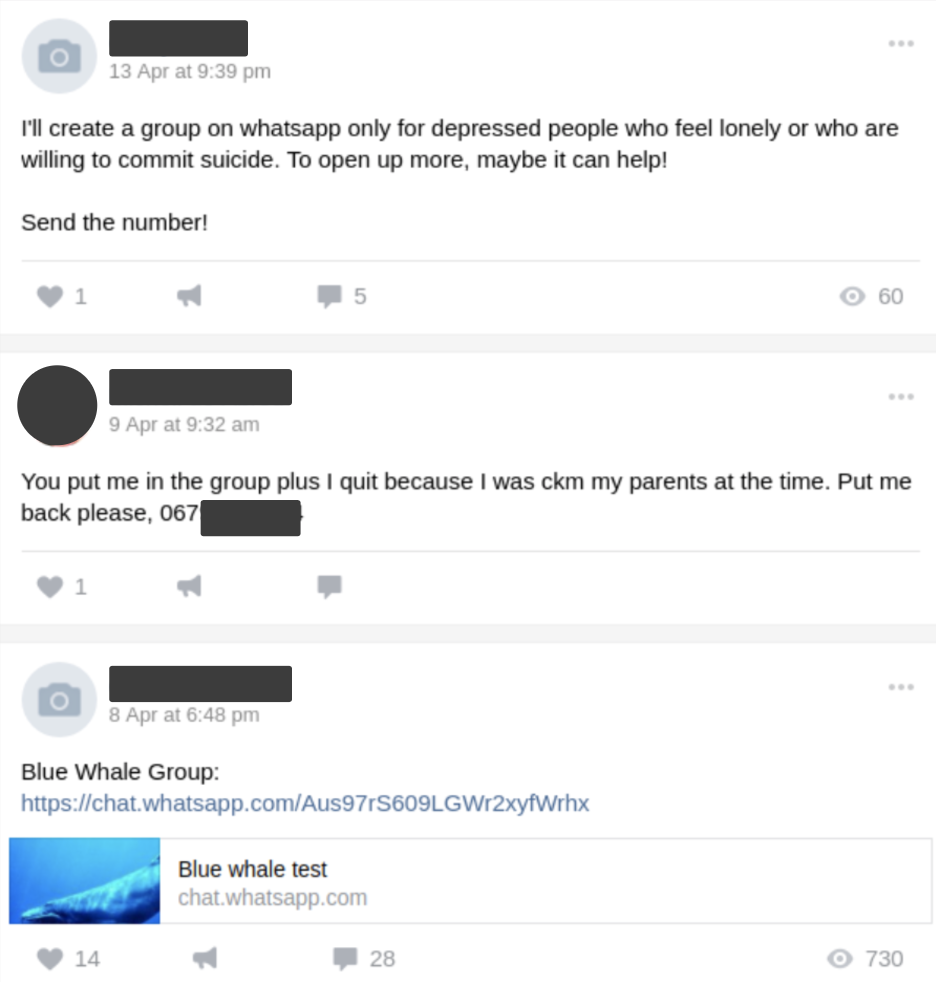}}}
\hfill\subfloat[A possible propagator.]{\label{main:b}\frame{\includegraphics[scale=.3]{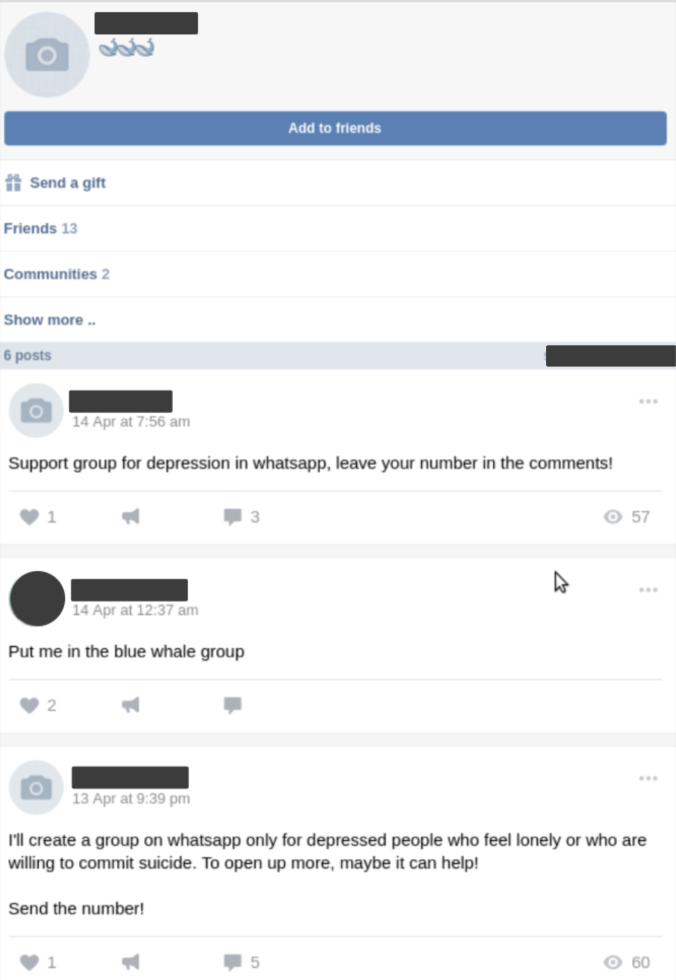}}} \\
\centering
\subfloat[Pictures of cut arms of users who are taking the Blue Whale Challenge. One or more tasks of the challenge involve conducting self mutilation and then taking pictures of the same.]{\label{main:c}\frame{\includegraphics[scale=.3]{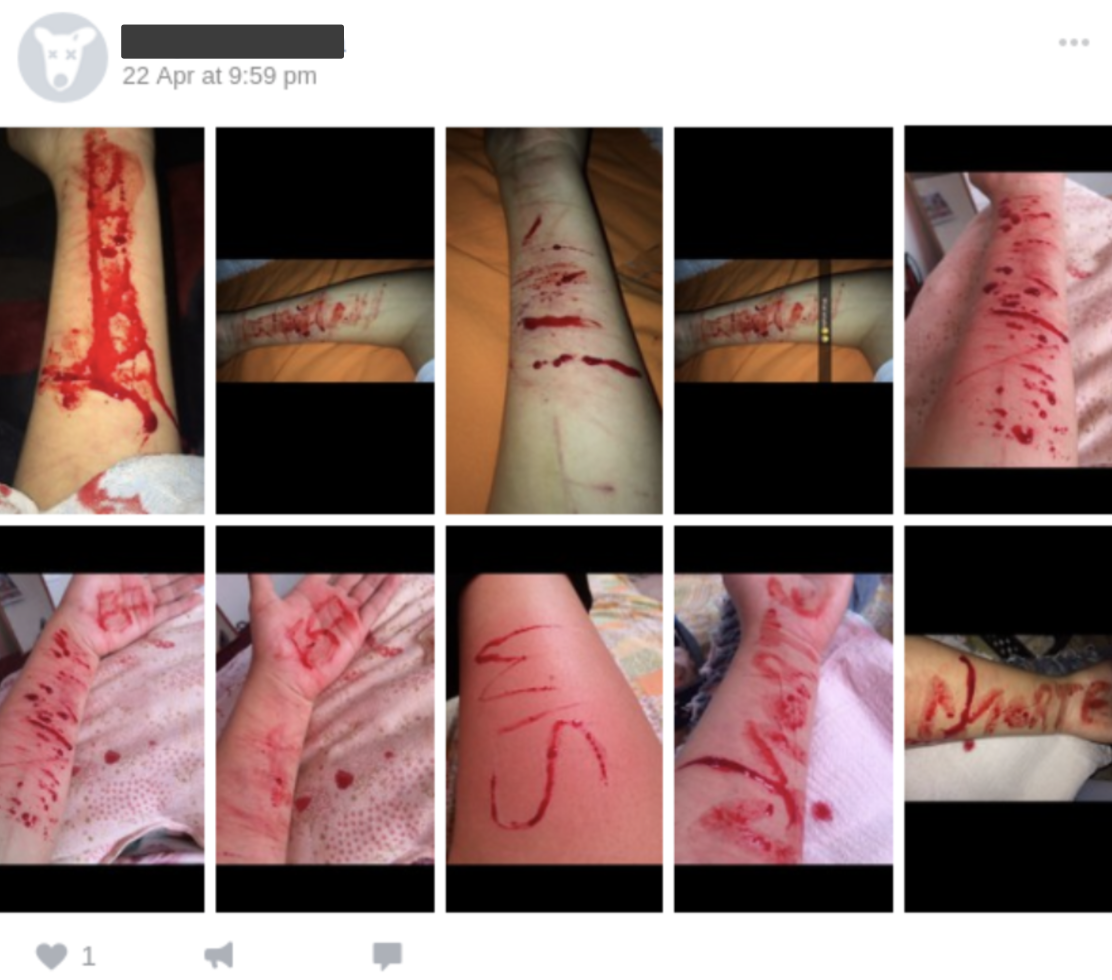}}}
\caption{VK - User Posts}
\label{vk_post}
\end{figure}

\subsection{Propagators and/or Pretentious Curator}
Users who post about the challenge with the intention of promoting it fall under this category. In extremely rare cases, it is possible that these propagators might be curators but such an event is counter-intuitive as actual curators would not risk revealing their identity. Fig. \ref{fig:curator} shows a Twitter account that claims to be a curator and asks users to private message or follow him/her to join the game. There have been cases where propagators or pretentious curators share the images of the 50 tasks (Fig. \ref{fig:task}) along with the links to APK files - which are not related to the game - misleading the users into believing that an application for the game exists. Some propagators also tend to share images of victims as shown in Fig. \ref{vk_post}(c). Fig. \ref{vk_post}(a) shows a propagator who shared a WhatsApp group link. Such things often excite users to reveal their personal information.
\begin{figure}
\centering
\subfloat[][Carve F57 - A task in the game.]{\frame{\includegraphics[width=0.55\textwidth,keepaspectratio]{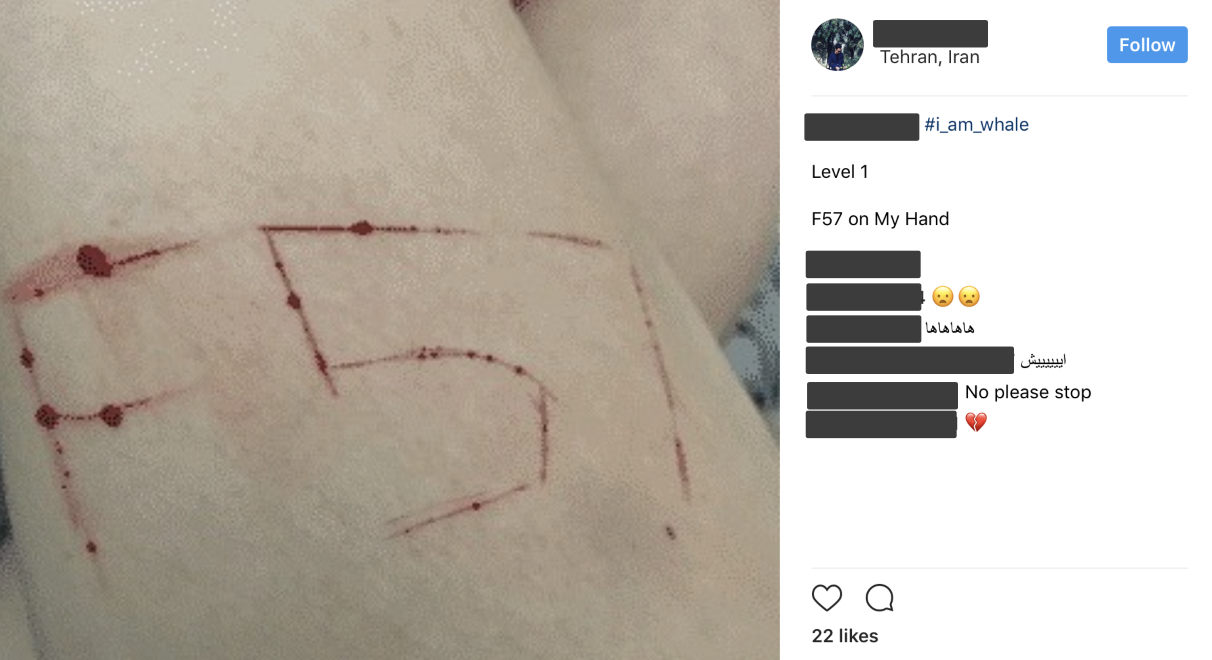}}}
\hfill
\subfloat[][Use of Blue Whale game related hashtags for popularity gain.]{\frame{\includegraphics[width=0.2\textwidth,keepaspectratio]{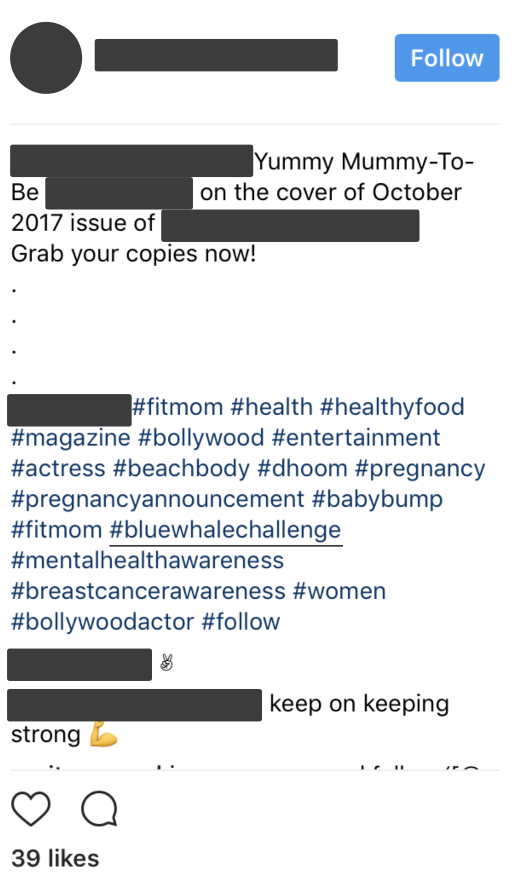}}}
\caption{Instagram - User Posts}
\label{insta_post}
\end{figure}

\subsection{Hashtag Hijackers}
There are users that use the Blue Whale challenge related hashtags in their posts just to seek attention and get some reactions to their posts. Fig. \ref{insta_post}(b) shows how irrelevant hashtags are used in order to garner more views and attention for the post. These people contribute to the noise in the data collected.

\section{Conclusion}
The Blue Whale challenge is a game that spread via online social media. It originated in Russia and is still on the rise; it is being propagated by the people themselves. On social media, people try using all types of keywords, hashtags, and images, so as to catch the attention of curators and be able to join the game. They might not even know what the game is about but want to play it. Then there are others who do not exclusively post only about the game but drop hints that they might be disturbed and looking for a way to end their life. A lot of sensitive information like phone numbers, email addresses etc. is revealed by people who want to take part in the challenge or those who are propagating it. A low fraction of people posting about the challenge follow up and post regularly. Users interested in Blue Whale Challenge are much better connected on VK than on Instagram. Also, the interaction between the users on VK - that is commenting on each other's posts - is much more than the interaction between such users on Instagram. The complexity of this game is that it is difficult to pinpoint which deaths are caused solely because of it. People may be depressed or affected by hardships before taking up the challenge. It is also possible that people who committed suicide showed the general symptoms that overlapped with those playing the game. Hence, it is difficult to verify deaths that are claimed to have occurred because of the challenge. Also, conversations between the curators and players are suspected to take place mainly through direct message - most of which are deleted like the posts on social media. Further, a lot of user accounts are deactivated or suspended.

\section{Actions taken by Social Media Services}

\begin{figure}[!htb]
\centering
\subfloat[][Instagram]{\frame{\includegraphics[width=0.3\textwidth,keepaspectratio]{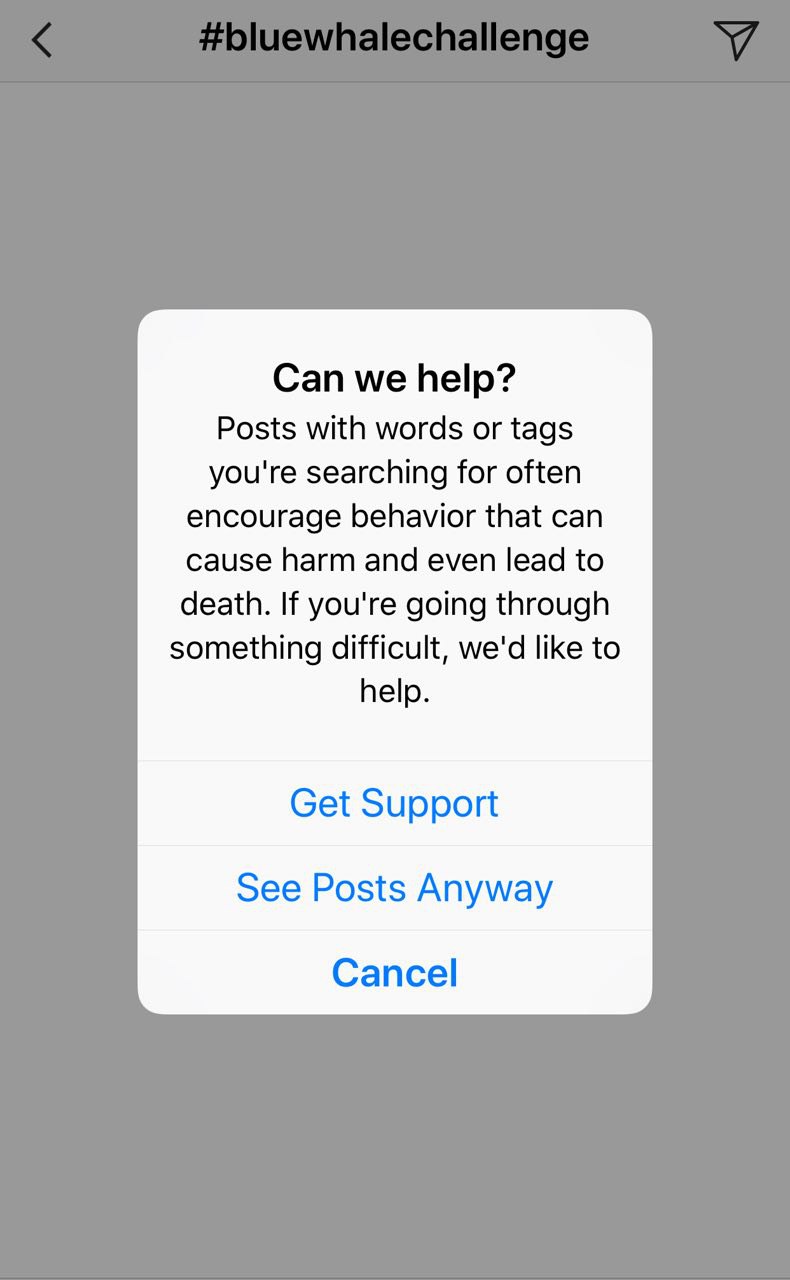}}}
\centering
\subfloat[][Tumblr]{\frame{\includegraphics[width=0.3\textwidth,keepaspectratio]{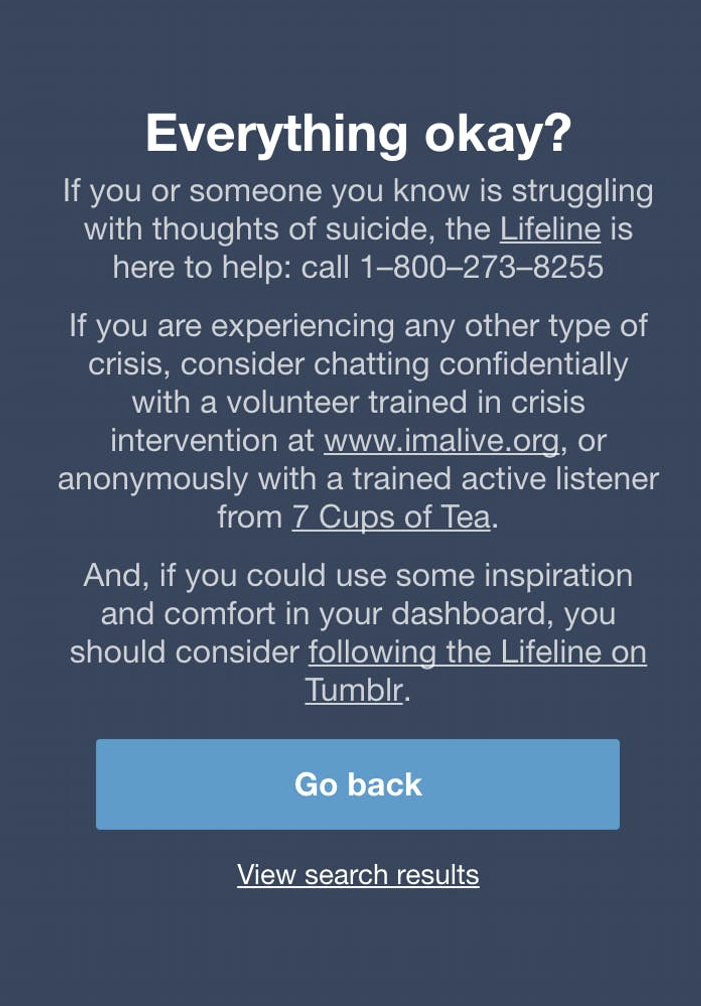}}}
\caption{Warning shown on searching about Blue Whale}
\label{warn}
\end{figure}

Instagram shows a warning when people search for pictures related to the Blue Whale Challenge. It offers help to people who might be going through something difficult but at the same time gives an option to \textquotedblleft see posts anyway\textquotedblright~\cite{12} as shown in Fig. \ref{warn}(a). Along with the warning, Tumblr also lists counselling and anti-suicide resources as shown in Fig. \ref{warn}(b). \\

In order to stop the spread of the challenge, a committee of experts has been set up by the government of India. The government has also asked companies like Google, Facebook, WhatsApp, Instagram, Microsoft and Yahoo to remove all links related to Blue Whale Challenge~\cite{22}. The supreme court has additionally asked major Indian News Channels to actively spread awareness about the Blue Whale Challenge~\cite{21}. The access of the Russian Social Network VK has also been temporarily banned in India~\cite{20}. 

\section{Limitations}
Most online social media websites have been instructed to remove posts pertaining to the Blue Whale Challenge. Images of cut hands and blood are often removed and suspicious user accounts are suspended. Also, a lot of users delete their posts. This leads to a shortage of data even though the effects of the game might be widespread. Also, API limits of social networks limit the amount of data that can be accessed in the first place. Due to lack of interaction with the individuals or those close to them, it is very difficult to know if there are other reasons why the victim is or was involved in the challenge.

\section{Future Work}
We would like to study the characteristics of vulnerable users and try to divide them into different categories like curator (difficult to find this), propagator, vulnerable player, beginner player etc. Based on this a confidence score can be developed which indicates the level of danger the user is in and what kind of interventions can be taken to get him/her out of the situation. If possible, we would like to do a detailed geographic analysis to find where most of these users are coming from. 
\section{Acknowledgements}
We would like to acknowledge the role of Srishti Gupta for providing her valuable inputs and suggestions throughout the project. We would also like to thank Kushagra Bhargava for getting us on track and evaluating our progress from time to time. Lastly, we would like to thank Vedant Nanda for his creative incites.
{\small
\bibliographystyle{ieee}
\bibliography{fenics}
}
\end{document}